\documentclass[12pt]{article}
\usepackage[utf8]{inputenc}
\usepackage{graphicx} 
\usepackage{dcolumn} 
\usepackage{bm} 
\usepackage{amsfonts}
\usepackage{amsthm}
\usepackage{amsmath}
\usepackage{amssymb}
\usepackage{epsfig}
\usepackage{color}
\usepackage{textcomp}
\usepackage{hyperref}
\usepackage{titlesec}
\usepackage{slashed}
\usepackage{caption}
\usepackage{subcaption}
\usepackage{siunitx}
\usepackage{booktabs}
\usepackage{multirow}
\usepackage{cite}
\usepackage{placeins}
\usepackage{comment}
\usepackage{braket}
\usepackage{authblk}

\let\a=\alpha \let\b=\beta  \let\g=\gamma  \let\d=\delta \let\e=\varepsilon
  \let\h=\eta    \let\k=\kappa \let\l=\lambda
\let\m=\mu    \let\n=\nu             \let\r=\rho
    \let\f=\phi 
   \let\o=\omega
   \let\L=\Lambda

\def\\{\hfill\break} \let\==\equiv

\let\0=\noindent

\def\nn{\nonumber}

\def\qed{\hfill\raise1pt\hbox{\vrule height5pt width5pt depth0pt}}

\def\be{\begin{equation}}
\def\ee{\end{equation}}
\def\bea{\begin{eqnarray}}\def\eea{\end{eqnarray}}

\begin{document}

\title{ \large \textbf{Graviton-Photon Oscillations in an Expanding Universe }}

\author[1]{Peter Anninos\thanks{anninos1@llnl.gov}}
\author[2]{Tony Rothman \thanks{tonyrothman@gmail.com}}
\author[3]{Andrea Palessandro\thanks{apalessandro@deloitte.dk}}
\affil[1]{\small Lawrence Livermore National Laboratory }
\affil[2]{\small New York University, Department of Applied Physics (retired)}
\affil[3]{\small Deloitte Consulting, Artificial Intelligence and Data}
\date{}

\maketitle

\begin{abstract}
Through the Gertsenshtein effect, the presence of a large external B-field may allow photons and gravitons to mix in a way that resembles neutrino oscillations and is even more similar to axion-photon mixing. Assuming a background B-field (or any field that behaves like one), we examine the Gertsenshtein mechanism in the context of  FLRW expanding universe models, as well as de Sitter space. The conformal invariance of Maxwell's equations and the conformal noninvariance of the Einstein equations preclude the operation of the Gertsenshtein effect at all scales.  In general we  find for the matter- and radiation-dominated cases, graviton-oscillations are possible only at late conformal times or when the wavelengths are much shorter than the horizon ($k\eta \gg 1$), but that the time-dependent oscillations eventually damp out in any case. The presence of charged particles additionally damps out the oscillations. For the de Sitter universe, we find that oscillations are possible only at early conformal times ($\eta \ll H^{-1}$) and for wavelengths short compared to the Hubble radius, but eventually freeze in when wavelengths become longer than the Hubble radius.  In principle a Gertsenshtein-like mechanism might influence the balance of particle species in an inflationary phase before freezing in; however, we find that in all our models the mixing length is larger than the Hubble radius.  We discuss several possible remedies to this situation.

\end{abstract}

\section{Introduction}
\setcounter{equation}{0}\label{intro}
\baselineskip 8mm

Sixty years ago Mikhail Gertsenshtein \cite{Gert61} demonstrated that the presence of a large external magnetic field can cause gravitational and electromagnetic waves to mix, producing oscillations that  in certain respects resemble neutrino-flavor oscillations. In a recent paper \cite{PR23a} Palessandro and Rothman gave a straightforward derivation of the Gertsenshtein mechanism, also pointing out that the axion-photon oscillations of current research interest are described by an essentially identical process.  In a companion paper \cite{PR23b} they showed that the self-interaction terms of the Yang-Mills SU(2) field can take the place of Gertsenshtein's external magnetic field to produce oscillations between gravitational waves and Yang-Mills ``bosonic waves." Both processes are classical, but in quantum language one can say that the external Maxwell or  the Yang-Mills magnetic field catalyzes the mixing of gravitons with photons or Yang-Mills bosons.  Such a description of course assumes that gravitons exist, a fact that has not, and perhaps never will be  experimentally established \cite{RB06, AP2020, Dyson05a, Dyson05b}. Yang-Mills SU(2) bosons are also entirely hypothetical.

The purpose of the present article is to investigate the behavior of the Gertsenshtein effect in expanding FLRW universes.  Several complications, however, make a consistent treatment of both gravitational waves and electromagnetic waves on an expanding background nontrivial. First, the usual d'Alembertian form of the gravitational wave equation found in textbooks gives incorrect results for nonstatic and curved spacetimes; one must therefore  solve the full Einstein equations directly.  Second, the introduction of an electromagnetic field with a preferred spatial direction breaks the symmetries of the isotropic and homogeneous FLRW universe; one must therefore either treat the waves as perturbations on an FLRW background or go to an anisotropic or inhomogeneous model. Furthermore, Maxwell's equations are known to be conformally invariant (see Wald \cite{Wald84}). Numerous authors \cite{Tsagas:2004kv,Jancewicz:2018zze, CR19, Cote19, Cot21}  have recently treated the consequences of this property in the context of cosmological electromagnetic fields. Its  importance for our investigation is that, while Maxwell's equations are conformally invariant, Einstein's equations are not.  Consequently, the symmetry between the gravitational wave equation and the electromagnetic wave equation is destroyed, making their solution more difficult, as well as immediately suggesting that the Gertsenshtein mechanism cannot operate at all scales.

In the following section we review the derivation of the Gertsenshtein effect given in \cite{PR23a}. As just implied, however, the comparatively simple formalism employed there cannot be consistently applied to an expanding cosmological model.  We must therefore adopt a more sophisticated covariant description for both the underlying cosmology and the electromagnetic field.  After introducing the new formalism in \S 2 and deriving the relevant wave equations, we then proceed to find solutions for the de Sitter universe (\S 3), radiation- and matter-dominated universe (\S 4) and curvature-dominated universes (\S 5).  For the latter cases, we  find damped mixing solutions at late times when particle wavelengths are much less than the Hubble radius\footnote{usually, but incorrectly, referred to as the``horizon."}, but that the solutions are frozen in at greater that Hubble-length scales, meaning that the Gertsenshtein effect does not function in that regime.  Surprisingly, for the de Sitter universe, the opposite occurs.  We find mixing solutions at early times, but not at late times.  Thus, it is possible that in an inflationary universe the Gertsenshtein mechanism might  operate for a time, until particle species are frozen out. Because the mixing length scale in today's universe is greater than the Hubble radius, the question arises as to whether it makes physical sense to study the Gertsenshtein mechanism in expanding models.  We discuss such issues in the conclusion and propose several reasons why such a study might be reasonable.\\

\section{Electrodynamics in an expanding universe}
\setcounter{equation}{0}\label{Maxwell}

As discussed above, oscillations between electromagnetic and gravitational plane waves can be made to take place when a large background magnetic field ${\bf B}_0$ couples the internal magnetic field of an electromagnetic wave  to the spacetime curvature produced by a gravitational wave traveling in the same direction.  The result is coherent oscillations between the  electromagnetic and gravitational wave; in particle language, the external magnetic field has catalyzed a resonant mixing between  photon and graviton states.  The derivation in \cite{PR23a} of this Gertsenshtein mechanism basically amounts to simultaneously solving the Maxwell-Einstein vacuum equations along with the gravitational wave equation, assuming that the source consists entirely  of the system's electromagnetic energy.

To summarize the procedure and introduce notation, for a flat, static spacetime we assume as usual that the gravitational wave (GW) is represented by a weak perturbation $h_{\m\n}({\bf x}, t) \ll 1$ traveling on a Minkowski background such that the full  metric is
\be
g_{\m\n} = \eta_{\m\n} + h_{\m\n} \ \ ;\ \  g^{\m\n} = \eta^{\m\n} - h^{\m\n}\ \ ;\ \   h^{\m\n} \equiv
\h^{\a\m}\h^{\b\m}h_{\a\b}, \label{gmunu}
\ee
where the flat-space metric is given by $\h_{\m\n} = (-1,1,1,1)$.  In the standard transverse-traceless (TT), or Lorenz, gauge, if the wave is assumed to be propagating in the $z$-direction, then $h_{\m\n} = h_{\m\n}(z,t)$ and the two independent polarizations of the GW are $h_{11} = -h_{22}$ and $h_{12} = h_{21}$.\footnote{We use units in which $G = c = 1$ and follow MTW \cite{MTW73} conventions throughout. In particular, Greek indices $= 0...3$ are spacetime indices, while Latin indices are spatial indices $= 1...3$. Repeated indices in any position are summed.}

In this situation the linearized gravitational wave equation is
\be\label{boxh}
\Box h_{\mu\nu} = -16\pi G T_{\mu \nu}.
\ee
and the stress-energy tensor is taken to be
\be
4\pi T_{ij}= -(E_iE_j + B_iB_j)+\frac1{2}\d_{ij}(\bf E \cdot E + B \cdot B)\label{Tij}
\ee
for total electric and magnetic fields of the system $\bf E$ and $\bf B$.  We assume that $\bf E$ is the electric field of the photon alone (small), while ${\bf B} = {\bf B}_0 + \bf b$ for external  field ${\bf B}_0$ and  photon magnetic field $\bf b$. 

The vacuum Einstein-Maxwell equations, which describe the propagation of the electromagnetic wave (EMW) on a curved background, can be written as
\be
\frac1{\sqrt{-g}}(\sqrt{-g} g^{\alpha \mu} g^{\beta \nu} F_{\m\n}),_\b = 0, \label{EMax}
\ee
where $F_{\m\n}$ is the usual flat-space electromagnetic field tensor. 

The trick in deriving the Gertsenshtein effect is to choose the external field ${\bf B}_0$ = {\bf constant} such that ${\bf B}_0 \perp \bf b$ and ${ B}_0 \gg b$.  This linearizes the equations; 
  for ${\bf b = b}_y$ and ${\bf B}_0 = {\bf B}_{0x}$ (\ref{boxh}) and (\ref{EMax}) give to first order in $h_{\m\n}$ and $b$ the coupled system
\bea
 \ddot b_y - b_y'' &=& h_{12}''B_{0},\label{EM7}\\
\ddot h_{12} - h_{12}'' &=& - 4B_0b_y.\label{boxh3}
\eea
(Here and below a dot ($\cdot$) represents the time derivative and a prime ($'$) indicates differentiation with respect to $z$.)

To establish graviton-photon mixing  we search  for ``wavepacket" solutions to these equations, i.e., we assume that both $b$ and $h$ (dropping subscripts) are product functions of a rapidly oscillating component and a slowly oscillating component:
\be
h= h_r(z-t)h_s(z+t) \quad ; \quad b= b_r(z-t)b_s(z+t),\label{hbansatz1}
\ee
 where ``r" stands for ``rapid" and ``s" stands for slow.  In the cases of interest, these expressions can be also written as 
 \be
 h(z,t) = \mathcal{A} e^{i\omega_r(z-\eta)} e^{-i \omega_s (z+\eta)}\quad ; \quad b(z,t) =\mathcal{B} e^{i\omega_r(z-\eta)} e^{-i \omega_s (z+\eta)},\label{hbansatz2}
 \ee
  with $\o_r \gg \o_s$.  The product decomposition allows us to find the beat frequency between the GW and the EMW, which is represented by the frequency of the slow component, $\o_s$.  One finds for the  system (\ref{EM7})-(\ref{boxh3})
\be
h_s = {\cal A}\sin\left(\frac{z+t+\phi}{L}\right)\ \  ;\ \ b_s = {\cal A} \sqrt{B_0} \cos\left(\frac{z+t+\phi}{L}\right),
\label{GEsol}
\ee
provided that $L \equiv 2/B_0$. Notice  that 
\be
h_s^2 + \frac{b_s^2}{B_0} ={\cal A}^2=\text{constant}, \label{A2}
\ee
which is a statement that energy is conserved under graviton-photon oscillations.

The quantity
\be
L  = \frac{2}{B_0} = \frac1{\o_s} \label{Lstat}
\ee
is the mixing length of the Gertsenshtein mechanism.  In cgs units 
\be
L = \frac{2 c^2}{\sqrt{G} B_0} \approx 2 \text{Mpc} \left(\frac{1 \,\text{Gauss}}{B_0}\right).
\ee

The strength of the intergalactic magnetic field is not known with any certainty, but $B_0 \approx 10^{-9}$ G appears to represent something like an upper limit (cf. \cite{Garr17}).    For this value of $B_0$  today's mixing length, $L_0$ is far greater than the Hubble radius of the universe, although around magnetars, where $B_0 \sim 10^{14}$ G, the mixing length can be only  $\sim 10^6$ km, making neutron stars  plausible graviton factories.\\

As already mentioned in the Introduction, given that $L_0$ is much greater than the Hubble radius, one can legitimately ask whether it makes  physical sense to consider  cosmological photon-graviton mixing.  We delay discussion of this question until \S\ref{Discussion}, after presentation of our results.  The immediate technical difficulty  confronting us is that due to the other issues discussed in the Introduction, the above  formalism, in particular the underlying equations (\ref{gmunu}) - (\ref{EMax}), cannot be consistently applied to the expanding universe scenario. We therefore adopt an approach similar to the one employed by Anninos \cite{Anninos98} for representing conformally decomposed matter-filled spacetimes and by Tsagas \cite{Tsagas:2004kv} for the elecromagnetic field description, both of which use a covariant 3 + 1 splitting of spacetime.

We write the general plane-symmetric, conformally decomposed metric in the form $g_{\mu\nu} = -dt^2 + \gamma_{ij} dx^i dx^j$ for both flat and non-flat topologies, with
\begin{equation}
\gamma_{ij} = \frac{a(t)^2}{{\cal R}(\overline{r})^2}  ~\widehat{\gamma}_{ij} 
            = \frac{a(t)^2}{{\cal R}(\overline{r})^2} \left[ \begin{array}{ccc}
                            a_1(t)(1-f(t,z)) & h(t,z)       & 0        \\
                            h(t,z)      & a_2(t)(1 +f(t,z)) & 0   \\
                            0           & 0                 & a_3(t)
\end{array} \right] ~,
\label{eqn:metricconformal}
\end{equation}where ${\cal R}(\overline{r}) = 1 + \kappa \overline{r}^2/4$, $\kappa = -1, ~0, ~+1$ for open, flat, and closed spacetimes respectively, $\overline{r}$ is related to the usual radius via $r=\overline{r}/(1+\kappa\overline{r}^2/4$), and $h(t,z)$ and $f(t,z)$ are first order (gravitational wave) perturbations. The $a_i(t)$ are constrained such that the determinant of the conformal metric $\widehat{\gamma}_{ij}$ is unity ($|\widehat{\gamma}|=1$), while $a(t)$ is the conformal factor which, for the case of matter dominated isotropic cosmologies takes the familiar form $a(t) \propto t^{2/3}$.

In order to simplify our analysis, however, we consider only cross-polarized gravity waves and set $f=0$ throughout. Further simplifications are made by adopting a geodesic slicing condition (unit lapse $\alpha=1$, zero shift $\beta^i$ in the 3+1 convention), isotropic background (constant and equal $a_i$), and (later) a conformal time coordinate $\eta=\int dt/a(t)$.

The total stress energy tensor is a linear combination of matter (perfect fluid plus cosmological constant) and electromagnetic fields $T^{\mu\nu} = T^{\mu\nu}_{\mbox{\small{m}}} + T^{\mu\nu}_{\mbox{\small{em}}}$, with
\begin{equation}
T^{\mu\nu}_{\mbox{m}} = \rho h ~u^\mu u^\nu + P_T g^{\mu\nu} \rightarrow \rho_E ~u^\mu u^\nu + P h^{\mu\nu} - \frac{\Lambda}{8\pi} g^{\mu\nu} ~,
\label{eqn:tmnHydro}
\end{equation}
and
\begin{align}
4\pi T^{\mu\nu}_{\mbox{em}} &= F^\mu_\lambda F^{\nu\lambda} - \frac{1}{4} g^{\mu\nu} F^{\lambda\kappa} F_{\lambda\kappa} \\
           &= \frac{1}{2} \left(E^\alpha E_\alpha + B^\alpha B_\alpha\right) \left(h^{\mu\nu} + u^\mu u^\nu\right)
               -\left(E^\mu E^\nu + B^\mu B^\nu\right) 
               + 2 u^{(\mu} \epsilon^{\nu)\alpha\beta} E_\alpha B_\beta ~,
\label{eqn:tmnEB}
\end{align}
where $F^{\mu\nu}$ is the Faraday tensor, $\Lambda$ is the cosmological constant, $\rho h = \rho + e + P$ is the proper relativistic enthalpy, $u^\mu$ is the covariant 4-velocity, and $h_{\mu\nu}=g_{\mu\nu}+u_\mu u_\nu$ is the projection operator orthogonal to the fluid frame. For simplicity we assume a barotropic equation of state $P = w \rho_E$ while absorbing the internal energy into a total energy density $\rho_E = \rho + e$. The barotropic coefficient  $w$ relates to the usual adiabatic index via $w=\gamma-1$. This form conveniently accommodates dust $w=0$ ($\gamma=1$), non-relativistic matter $w=2/3$ ($\gamma=5/3$), and relativistic matter or radiation $w=1/3$ ($\gamma=4/3$). 

For the electromagnetic field we follow the convention in \cite{PR23a, PR23b} and define weak (relative to the background energy density) electromagnetic field vectors of the form
\begin{align}
E^i &= a(t)^{-3} [e(t,z), ~0     , ~0 ] ~,   \label{eqn:pertE} \\
B^i &= a(t)^{-3} [B_0, ~b(t,z) ,~0] \label{eqn:pertB}
\end{align}
where $B_0$ is the constant value of the external magnetic field, and the factor $a(t)^{-3}$ is introduced to scale away the metric determinant and simplify the resulting differential equations. Perturbation equations for the metric and fields are derived assuming $B_0^2 \ll \rho_E a^4$ to preserve isotropy at zero order. Formally we have introduced two smallness parameters defining  the field strength relative to the dominant cosmological energy density $\epsilon \equiv B^\mu B_\mu/\rho_E \ll 1$, and the gravitational wave amplitude. We point out the electromagnetic energy density $B^\mu B_\mu$ scales to lowest order as $\propto B_0^2 a^{-4}$, similar to $\rho_E$ as expected for radiation-like behavior consistent with magnetic flux conservation.\footnote{Note for $B_0=0$ in \ref{eqn:pertB}, $B^2 = B_i B^i \equiv a^2 B^iB^i = a^{-4} b^2$, implying that adiabatic decay of $B^2 \propto a^{-4}$ is satisfied by $b=$ constant, and therefore $b$ is interpreted as the magnetic flux.\label{footnote b}}  

It is not possible to derive plane wave equations in non-flat spacetimes except in the limit of small (localized) spatial volumes such that $\overline{r} \ll 1$.  Moreover, in order to pick up all contributions to the field equations from  curvature it is important to keep spatial derivatives  throughout the complete derivation of the second order PDEs due to the 3D nature of the metric and use of Cartesian coordinates. For this purpose we evaluate the following second order differential representation of Maxwell's equations
\begin{equation}
\Box{F^{ab}} + g^{be} \nabla_d\nabla_e F^{da} + g^{ae} \nabla_d \nabla_e F^{bd} = 0 ~,
\end{equation}
before taking the limit $(x,y,z) \rightarrow 0$.

In the $\overline{r} \ll 1$ limit we derive  general differential equations applicable to all the spacetimes considered in this paper.  For the GW:
\begin{equation}
\ddot{h} - \frac{h''}{a^2} + 3\frac{\dot{a}}{a} \dot{h} - \frac{\kappa}{a^2} h + \frac{4}{a^4} B_0~b = 0.
\label{eqn:pert_ddoth_nonflat}
\end{equation}
For the EMW:
\begin{equation}
\ddot{b} - \frac{b''}{a^2} + \frac{\dot{a}}{a} \dot{b} - \frac{B_0}{a^2} ~h'' +\frac{\kappa}{a^2} \left( 2 b + B_0~h \right)= 0.
\label{eqn:pert_ddotb_nonflat}
\end{equation}
For  the background (isotropic) expansion:
\begin{equation}
\frac{\dot{a}^2}{a^2} - \frac{\Lambda}{3} - \frac{8\pi \rho_0}{3 a^{3(1+w)}} + \frac{\kappa}{a^2}  = 0 ~,
\end{equation}
where $\rho_0$ is a constant defining the total energy density of matter or radiation. Notice we have neglected electromagnetic field energy from the background expansion as a higher order contribution, which would otherwise have  appeared as $\propto B_0^2/a^4$. Also, (\ref{eqn:pert_ddotb_nonflat}) reduces to that in \cite{Tsagas:2004kv} when $B_0 = 0$. 

The above equations  simplify considerably when rewritten in conformal time $\eta \equiv \int dt/a(t)$, giving
\begin{equation}
\ddot{h} - h'' + 2\frac{\dot{a}}{a} \dot{h} - \kappa h + \frac{4}{a^2} B_0~b = 0 ~,
\label{eqn:nonflat_h}
\end{equation}
\begin{equation}
\ddot{b} - b'' - B_0 ~h'' + \kappa \left( 2 b + B_0~h \right) = 0 ~,
\label{eqn:nonflat_b}
\end{equation}
\begin{equation}
\frac{\dot{a}^2}{a^2} - \frac{\Lambda a^2}{3} - \frac{8\pi \rho_0}{3 a^{1+3w}} - 
\frac{B_0^2}{3a^2} + \kappa = 0 ~,
\label{eqn:friedman}
\end{equation}
where for future reference (\S\ref{Discussion}) we have now included the (assumed small) magnetic field energy density in the Friedmann equation.

The most obvious feature of the above is that the GW equation contains a damping term, while the $B$-field equation does not.  This is a reflection of the conformal invariance of Maxwell's equations and will be responsible for much of the behavior of the solutions.

We also point out that to close equations (\ref{eqn:nonflat_h}) - (\ref{eqn:friedman}) we need an explicit expression for the scale factor as a function of (conformal) time, essentially a solution to the Friedmann equation (\ref{eqn:friedman}). In general this is not trivial, and so in order to retain these equations in manageable form, we consider only solutions in regimes where the scale factor is approximated by a simple power law or exponential. In the following sections, we discuss each of these standard cases, finding analytic solutions at both early and late cosmological times.

\section{de Sitter space}
\label{desitter}

Here we examine the flat ($\k = 0$), exponentially expanding case $a = e^{Ht}$, which written in conformal time is
\be
a(\eta) = (1-H\eta)^{-1}.
\ee
Note that $H^2 = \L/3$ for cosmological constant $\L$ and that from (\ref{eqn:friedman}) with $\r_B \sim B^2 \propto a^{-4}$, we are tacitly assuming the constraint $B_0^2/a^4 \ll \L$.  In the de Sitter model, equations  (\ref{eqn:nonflat_h}) and (\ref{eqn:nonflat_b}) become 
\begin{equation}\label{bheqL}
    \begin{split}
     &\ddot h - h'' + 2 \frac{H}{(1-H\h)}\dot h + {4 \left({1-H\h}\right)^2   B_0 b} = 0,\\
     &\ddot b - b'' - B_0 h''= 0.
    \end{split}
\end{equation}

This system evidently has analytic solutions for both early and late times. For early times $H\eta \ll 1$, and so we have simply
\begin{equation}\label{bheqLe}
    \begin{split}
        &\ddot h - h'' + 2(aH)\dot h + {4B_0 a^{-2}  b} = 0,\\
        &\ddot b - b'' - B_0 h''= 0.  
    \end{split}
\end{equation}

(Although $a \approx 1$ at early times we have explicitly included it here in order that the mixing length $L$   be expressed as a function of the scale factor; see e.g. \ref{Lprop}.) Now following the procedure of \cite{PR23a} and \cite{PR23b} we  assume a wave packet solution of the form (\ref{hbansatz1})-(\ref{hbansatz2}), where the rapidly varying components $b_r$ and $h_r$ describe the propagation of electromagnetic and gravitational wave packets of energy $\omega_r$, while the slowly varying components $b_s$ and $h_s$ describe the coherent oscillations between them with frequency $\omega_s \ll \omega_r$. However, since (\ref{bheqLe}) for $h$ includes a damping term, we  anticipate that its solution should contain a factor $e^{-\g \h}$ with  damping constant $\g$ and modify the ansatz slightly to read 
\begin{equation}\label{dSansatz}
\begin{split}
h (z,\eta) &= \mathcal{A} e^{i\omega_r(z-\eta)} e^{-i \omega_s (z+\eta)}e^{-\g\h},\\
b (z,\eta) &= \mathcal{B} e^{i\omega_r(z-\eta)} e^{-i \omega_s (z+\eta)}.
\end{split}
\end{equation}

Substituting the expression for $h$ into the first of (\ref{bheqLe}) gives
\be
\mathcal{A} \left[-4\o_r\o_s + 2i\g(\o_r+\o_s) - 2iaH(\o_r+\o_s)+ \g^2 - 2 a H\g \right] e^{-\g \h}  = -4\mathcal{B} B_0 a^{-2}.  \label{hans}
\ee
The imaginary part must vanish,  implying that $\g = aH$. Also, $H \eta \ll 1$, so we can assume $e^{H \eta} \approx 1$. Thus
\be
\frac{\mathcal{A}}{\mathcal{B}} = \frac{4 B_0 a^{-2}}{4 \o_r\o_s + (aH)^2}.
\ee

Substituting the anzatz for $b$ into the second of (\ref{bheqLe}) gives
\be
4\mathcal{B}\o_r\o_s = \mathcal{A}B_0 (\o_r - \o_s)^2 \approx \mathcal{A}B_0 \o_r^2,  \label{bans}
\ee
or
\be
\frac{\mathcal{A}}{\mathcal{B}} = \frac{4\o_s}{B_0\o_r}.
\ee
Eliminating $\mathcal{A}/\mathcal{B}$ from the above two equations yields a quadratic equation with solutions
\be\label{osdesitter}
\o_s = -\frac{(aH)^2}{8 \omega_r} \left[1 \pm \sqrt{1+ 16 \frac{  B_0^2 \omega_r^2 a^{-2}}{(aH)^4}} \right].
\ee
Thus, contrary to the static case, in de Sitter the mixing frequency depends on  $\omega_r$, as well as on $H$. The overall sign in (\ref{osdesitter}) is not important and can be changed by choosing $\o_s$ positive in ({\ref{dSansatz}), which does not alter the underlying interpretation that $\o_s$ is the beat frequency produced by the superposition of two waves traveling in the same direction. The choice of sign for the radical reflects  two different  solutions for the decoupled GW and EMW equations in the limit $B_0=0$. When $B_0 = 0$, equation (\ref{hans}) is satisfied for (non-zero) GWs $h$ by choosing the positive radical, which results in $\o_s = -(aH)^2/4\o_r$; the negative radical gives $\o_s=0$, which is appropriate for linearized plane wave solutions of equation (\ref{bans}) for $b$.

Despite that for $B_0 = 0$ the GW solution is $\o_s = (aH)^2/4\o_r$, from (\ref{eqn:nonflat_h}) and \ref{eqn:nonflat_b} it is evident that in this limit  for  FLRW models the coupling between $b$ and $h$ vanishes.  Thus no graviton-photon mixing  can be taking place. The GW and EMW oscillate independently; the beat frequency $\o_s$ in the metric perturbations evidently arises through a natural ``resonance" with the background expansion that dampens gravitational waves at the Hubble rate, and which vanishes as $H\rightarrow 0$. 

Note that in the limit  $B_0 \omega_r \gg (aH)^2$, (\ref{osdesitter}) gives for either sign option
\begin{equation}
    L \equiv \frac{1}{\omega_s} =  \frac{2a}{B_0} ~,\label{Lprop}
\end{equation}
which reduces to the static result (\ref{Lstat}) for $a\rightarrow 1$, as expected.

Although we have thus far  made only an early time approximation, the damping of $h$ strongly suggests that the Gertsenshtein mechanism cannot function  forever in the de Sitter universe; either  the gravitational waves effectively disappear or, as is known \cite{RA97}, they freeze-out when their wavelengths reach superhorizon scales, causing mixing to cease.  This suspicion is  confirmed by  examining the late-time behavior of Eqs. (\ref{bheqL}).

To find a late-time solution, let $\e \equiv (1 - H\h)$.  Then  
\begin{equation}\label{desitter_late}
    \begin{split}
        &\ddot {h} - \frac{2}{\e}\dot h - 
        \frac{h''}{H^2}  = - \frac{4   B_0 b}{H^2}\e ^2,\\
        &H^2\ddot b - b'' = B_0 h'', 
    \end{split}
\end{equation}
where ($\cdot$) now indicates $d/d\e$.  We assume a separable solution $h = h(\e) h(z)$.  The equation for  $h(z)$ is simply
\be
h''(z) = -\hat{k}^2H^2 h(z),
\ee
where $\hat{k}^2$ is the (dimensionless) separation constant.  The above of course implies oscillatory behavior $h(z) \sim e^{-i\hat{k}H z}$.  For the homogeneous ($B_0 = 0$) case, the equation for $h(\e)$ becomes
\be\label{bessel}
\ddot h(\e) - \frac{2}{\e}\dot h(\e) + \hat{k}^2 h(\e) = 0,
\ee
which is Bessel's equation with solutions
\be
h(\e) = \e^{+ 3/2}J(\hat{k}\e)_{\pm 3/2}.
\ee

Since the late-time behavior of the system corresponds to $\e \to 0$, we can expand the Bessel functions in powers of $\e$ and retain the first few terms. To order $\e^2$ 
\be
h(\e) = h_0 + \frac{\hat{k}^2}{2}h_0\e^2 
\ee
satisfies the homogeneous equation ($B_0 = 0$) to lowest order. If for the inhomogeneous case we  keep the same form for $h(\e)$ and additionally expand  $b$  as
\be
b(\e) = b_0 + b_1\e^2,
\ee
with the same spatial dependence as $h(z)$, we find upon substituting into (\ref{desitter_late}) that the  solutions
\begin{align} 
b_0 &= -\frac{\hat{k}^4 H^2 h_0}{8 B_0} ~, \\
b_1 &= -\frac{B_0 \hat{k}^2 h_0}{2} ~,
\end{align}
satisfy both the GW and EMW equations to order $H^2 \hat{k}^4/B_0^2 \ll 1$ in the long wavelength limit $k\equiv \hat{k}H \ll a H$; here we have had to additionally impose the condition that $B_0 \ll \sqrt{\Lambda} a^2 = H a^2$ for de Sitter (see (\ref{eqn:friedman})).

For a less restrictive condition on the wavelength, the ansatz 
\be
h = h_0 + B_0 c_1 + \left(\frac{\hat{k}h_0}{2} + \frac{B_0 c_2}{2}\right)\e^2,
\ee
which reduces to the previous when $B_0= 0$, leads to $h = b= 0$ to order $\e^2$, which of course precludes any oscillations, including waves of the type (\ref{dSansatz}).\footnote{Note that solutions (\ref{dSansatz}) are separable; we  merely wrote them in a way to emphasize their wavelike character.} 

We therefore conclude that the Gertsenshtein effect is inoperable in de Sitter space at late times $H\eta \lessapprox 1$.

\section{Radiation- and matter-dominated universes }
\label{MDRD}

For the RD and MD universes, equations (\ref{eqn:nonflat_h}) and (\ref{eqn:nonflat_b}) become
\begin{equation}\label{bheqn}
    \begin{split}
        &\ddot h  - h'' + 2  \frac{p}{\h} \dot h  -\kappa h= - \frac{4 B_0 b}
    {(\eta/\eta_0)^{2p}} ,\\
        &\ddot b - b'' + 2\kappa b = B_0 (h'' - \kappa h) , 
    \end{split}
\end{equation}
where $(\eta/\eta_0)^p = a$ and $p = 1,2$ for the RD  and MD  cases, respectively.  Here, we must assume that $|B|^2 \ll \rho_E$, or equivalently $B^2_0 \ll \rho_E a^4$.  Additionally, since we retain curvature in these equations we must also have $\kappa \ll \rho_E a^2$, but leave open for the moment the relative scaling between magnetic field strength and curvature.

As in the late time de Sitter case, we first solve these equations by separation of variables.  For $h= h(\h)h(z)$ with separation constant $k$, we have for the flat-space decoupled situation 
\be\label{bessel2}
\ddot h(\h) + 2  \frac{p}{\h} \dot h  + k^2 h(\h) = 0 \quad ; \quad h''(z) = -k^2h(z).
\ee
The  former is again a Bessel equation, and so the full (decoupled) inhomogeneous solutions are
\be\label{RDhom}
\mathrm{RD}:\  h_o(z,\h) \propto\h^{-1/2}J_{\pm 1/2}(k\h)e^{-ikz} \qquad; \qquad \mathrm{MD}:\  h_o(z,\h) \propto \h^{-3/2}J_{\pm 3/2}(k\h)e^{-ikz}
\ee

Guided by these solutions, we now attempt to solve the inhomogeneous equations for the RD and MD cases in the limits $k\h \ll 1$ and $k\h \gg 1$.

\subsection{Radiation-dominated, $k\h \ll 1$}

 This represents solutions for which $\l \gg \h \sim H^{-1}$, in other words ``superhorizon" (super--Hubble-radius) evolution.  Since we are also restricted to $\overline{r} \ll 1$, it also represents very early times.  The asymptotic form of the $J_{1/2}$ Bessel function for small argument suggests we try solutions of the form
\bea
h &=& [h_0 + h_1 (k\h)^2 + h_2 (k\h)^4]e^{-ikz} ~,\nn\\
b &=& [b_1(k\h)^2 + b_2(k\h)^4]e^{-ikz} ~.
\eea 
Inserting these expressions into (\ref{bheqn}) and equating powers of $\h$ leads to four equations for the  five constants.  After some algebra we find
\bea 
b_1 &=& -\frac{B_0h_0}{2} ~(1+\overline\kappa) ~, \nn\\
h_1 &=& \frac{b_1}{3B_0}\left (\frac{1-\overline\kappa}{1+\overline\kappa} - 2 B_0^2\eta_0^2 \right) ~, \nn\\
b_2 &=& \frac{b_1}{18}\left(B_0^2 \eta_0^2 (1+\overline\kappa) - \frac52\overline\kappa - 2\right) ~, \nn\\
h_2 &=& -\frac{b_1}{60B_0} \left( \frac{(1-\overline\kappa)^2}{1+\overline\kappa} + \frac13 B_0^2 \eta_0^2(\overline\kappa - 10) + \frac23 B_0^4\eta_0^4 (1+\overline\kappa)\right) ~,
\eea 
where $\overline\kappa=\kappa/k^2$, and $h_0$ is a freely specifiable parameter. These solutions satisfy both field and gravity wave equations to order $(k\eta)^4$, and have the identical power-law behavior as in the decoupled ($B_0=0$) case.

We note that these solutions are compatible with those of Tsagas\cite{Tsagas:2004kv} and Grishchuk\cite{Grish74} for the ``superadiabatic" amplification of EMWs and GWs, meaning that the magnetic flux $b$ increases over time{\footnote{See footnote \ref{footnote b}} and $h$ falls off more slowly than $a^{-1}$ at ``superhorizon" scales.\footnote{$\r_h \propto \dot h^2 \propto \o^2 h^2$.  Both $\o$ and $h$ decrease as $a^{-1}$, so $\r_h\propto a^{-4}$.}   The exact time dependence of $b$ for the growing modes considered here differs  from Tsagas's, given that he assumed an exponential solution $a \sim e^\h$ and we have assumed a power law $a \sim \h^p$.  Our time dependence for $h$ is exactly that of Grishchuk for both the RD and MD cases (cf. \S\ref{MDsub}, below).

For us the important point is that, because these solutions are not oscillatory, they once again  preclude the Gertsenshtein effect from operating in this regime.

\subsection{Radiation-dominated, $k\h \gg 1$}\label{subRD}

This represents ``subhorizon" (sub--Hubble-radius) late-time behavior. 

We directly attempt to find a mixing solution.  Analogously to de Sitter  we assume
\begin{equation}\label{ansatz1}
\begin{split}
b (z,\eta) &=  \mathcal{B} e^{i\omega_r(z-\eta)} e^{-i \omega_s (z+\eta)}(k\eta)^n, \\
h (z,\eta) &= \mathcal{A} e^{i\omega_r(z-\eta)} e^{-i \omega_s (z+\eta)}(k\eta)^m.
\end{split}
\end{equation}
Equation (\ref{bheqn}) for $h$ then gives 
\be
-4\o_r\o_s -\frac{2i(m+1)}{\h}~(\o_r + \o_s) + \frac{m(m+1)}{\h^2} - \kappa = -\frac{\cal B}{\cal A} 4 B_0 k^2 \eta_0^2 (k\h)^{n-m- 2}.  \label{RDh1}
\ee
Note that choosing  $m=-1$ causes the imaginary part and higher-order terms on the left to vanish, resulting in the correct asymptotic form for the decoupled solution (\ref{RDhom}). Similarly, (\ref{bheqn}) for $b$ gives
\be
 -4\o_r\o_s -i \frac{2n}{\h}(\o_r + \o_s) + \frac{n(n-1)}{\h^2} + 2\kappa = -B_0((\o_r-\o_s)^2 + \kappa) ~\frac{\mathcal A}{\mathcal B} (k\h)^{m-n}. \label{RDb1}
\ee
Here, choosing $n=0$ causes the imaginary part to vanish, meaning that $b$ is undamped.  Now, with $(\o_r - \o_s)^2 \approx \o_r^2$, (\ref{RDb1}) becomes 
\be
 -4\o_r\o_s - i \frac{2n}{\h}(\o_r + \o_s) + \frac{n(n-1)}{\h^2} +  2\kappa = -B_0(\o_r^2 + \kappa) ~\frac{\mathcal A}{\mathcal B} (k\h)^{m-n} ~, \label{RDb1a}
\ee
or
\be
\frac{\mathcal B}{\mathcal A} = -\frac{(k\h)^{m-n}B_0 (\o_r^2+\k)}{-4\o_r\o_s - \frac{2in(\o_r +\o_s)}{\h} + \frac{n(n-1)}{\h^2}+ 2\k}.
\ee
Inserting this into (\ref{RDh1}) yields, for any $n,m$,
\be
\begin{split}
\left(-4\o_r\o_s -\frac{2i(m+1)}{\h}~(\o_r + \o_s) + \frac{m(m+1)}{\h^2} - \kappa\right)\\
\times \left(-4\o_r\o_s - \frac{2in(\o_r +\o_s)}{\h} +  \frac{n(n-1)}{\h^2} + 2\k\right) 
&= \frac{4 B_0^2 \eta_0^2 k^2(\o_r^2 + \kappa) }{(k\h)^{ 2}}.  \label{RDbh}
\end{split}
\ee
Notice that the RHS is independent of $n,m$.  Thus, since $k\h \gg 1$, if we neglect  all terms with powers of $k\h$ in the denominators for a lowest-order approximation, we can equate either factor in parentheses to zero, which is the same as setting the LHS of  (\ref{RDh1}) or (\ref{RDb1}) to zero with the higher-order terms neglected.  The result is:
\be
\o_s = \frac{\k}{2\o_r}\ , \ \frac{-\k}{4\o_r}. \label{wk}
\ee
Since the coupling with $B_0$ has been neglected, this evidently represents a ``curvature-induced" resonance similar to the effect discussed in \S\ref{desitter}, but it does not represent graviton-photon mixing.

If we now restrict ourselves to the physically motivated solution $m = -1$, then in (\ref{RDbh}) we must have $n= 0$ to kill the imaginary part and retain a non-negligible coupling term.  This results in the quadratic
\be
8\o_r^2\o_s^2 - 2\o_r\o_s\k - \k^2 - \frac{2B_0\eta_0^2(\o_r^2 + \k)}{\h^2} = 0,
\ee
with solution
\be
\o_s = \frac{\k}{8\o_r}\left[1 \pm \sqrt{1 + 8 \left[1+ \frac{2B_0^2\eta_0^2(\o_r^2+\k)}{\k^2 \h^2}\right]}\right].  \label{o_sRDsol}
\ee
Notice that if the last term is neglected then the roots are the same as the frequencies  (\ref{wk}).  On the other hand, in  the high frequency limit  ($\o_r^2 \gg \k$) the result simplifies to 
\be
\omega_s = \frac{B_0}{2 a} ~,
\ee
or equivalently
\be
 L = \frac{1}{\omega_s} = \frac{2a}{B_0} ~,
\ee
which  again recovers the static solution for the mixing length.

Since $n=0$ in this case, the amplitude of $b$ remains constant.  We interpret this to be a result of the conformal invariance of the electromagnetic field discussed earlier. Regardless of the choice of $m$ and $n$, however, the  result suggests that in flat space (with zero or negligible curvature) the mixing frequency decreases as the universe expands until there is no mixing whatsoever and the Gertsenshtein effect ceases to operate.

We make a final point regarding the curvature terms in the above equations. In deriving these solutions we have implicitly assumed that radiation (not curvature) determines the background cosmology. The Friedmann equation
\begin{equation}
    \dot{a}^2 = \frac{8\pi \rho_0}{3 a^{1+3w}} - \kappa 
    = \frac{8\pi \rho_0\eta_0^2}{3 \eta^2} - \kappa~
\end{equation}
puts a relative scale on the curvature terms, which is valid for both RD ($w=1/3$) and MD ($w=0$) models. To satisfy radiation-dominance (or dust-dominance, next section) with $\k = \pm 1$, we require $8\pi \rho_0\eta_0^2/3\eta^2 \gg 1$.  Together with the subhorizon condition $(k\eta) \gg 1$, the  solutions in this section are therefore valid only for $1\ll (k\eta)^2 \ll 8\pi\rho_0 k^2/3$, effectively putting a time limit on their validity.  Beyond this limit we must consider solutions in a curvature-dominated regime, which we do  in \S\ref{CD}.\\

\subsection{Matter-dominated, $k\h \ll 1$}\label{MDsub}

This represents early-time, ``superhorizon" (super--Hubble-radius) solutions.  

The asymptotic form of $J_{3/2}$ in (\ref{RDhom}) for small argument suggests that for the coupled case we attempt a solution of the form
\bea
h(\h) &=& h_0 + h_1(k\h)^2 + h_2(k\h)^4 ~, \nonumber\\
b(\h) &=& b_0 + b_1(k\h)^4 + b_2(k\h)^6 ~.
\eea
Inserting this ansatz into (\ref{bheqn}) and defining $\overline\kappa=\kappa/k^2$ as we did earlier, gives $h_0=b_0=0$ along with
\bea
h_1 &=& -\frac{12 b_1}{B_0(1+\overline\kappa)} ~, \nonumber\\
h_2 &=& \frac{b_1}{6B_0} \frac{4-\overline\kappa}{1+\overline\kappa} ~, \nonumber\\
b_2 &=& -\frac{b_1}{180} (10 + 11\overline\kappa) ~,
\eea
where $b_1$ is arbitrary, but because $h_0 = 0$ one finds two conditions relating $h_1$ and $b_1$, which result in the constraint
\begin{equation}
B_0^2 k^2 \eta_0^4 (1+\overline\kappa) = 30 ~.
\end{equation}
Again, since these solutions are frozen-in, rather than oscillatory, the Gertsenshtein effect is evidently inoperable in this regime. 

\subsection{Matter-dominated, $k\h \gg 1$}
This represents late-time, ``subhorizon" (sub--Hubble-radius) solutions.

We follow the identical procedure as for the RD $\k\h \gg 1$ case.  The only difference is that in (\ref{bheqn}) $p = 2$.  Assuming ansatz (\ref{ansatz1}) for $b$ and $h$, the GW equation becomes
\be
-4\o_r\o_s -\frac{2i(m+2)}{\h}~(\o_r + \o_s) + \frac{m(m+3)}{\h^2} - \kappa = -\frac{\cal B}{\cal A} 4 B_0 k^4\eta_0^4 (k\h)^{n-m- 4}.  \label{MDh1}
\ee
Note that in this case $m = -2$ causes the imaginary part to vanish and gives the correct asymptotic behavior for the decoupled solution.  Eq. (\ref{RDb1}) for $b$ is unchanged, and so eliminating ${\cal B/A}$ as before gives for $m=-2$
\be
\left(-4\o_r\o_s - \frac{2}{\h^2} - \kappa\right)
\left(-4\o_r\o_s - i\frac{2n(\o_r +\o_s)}{\h} +  \frac{n(n-1)}{\h^2} + 2\k\right) 
= \frac{4 B_0^2 k^4 \eta_0^4 (\o_r^2 + \kappa) }{(k\h)^{ 4}}~,  \label{MDbh}
\ee
where we have also set $(\omega_r-\omega_s)^2 \approx \omega_r^2$.
The lowest-order constant solutions found by ignoring all higher-order terms $(k\eta)^{-1}$ and smaller (including the coupling term), are the same as in the RD case:
\be
\o_s = \frac{\k}{2\o_r}\ , \ \frac{-\k}{4\o_r}. 
\label{wk1}
\ee

If we additionally set $n=0$, the remaining imaginary term vanishes, as well as the $n(n-1)$ term, and the coupling term can be retained to derive a time-dependent mixing frequency. In this case the resulting quadratic becomes
\be
8\o_r^2\o_s^2 + \omega_s\left(\frac{4\omega_r}{\eta^2} - 2\o_r\k\right) - \k^2 -\frac{2\kappa}{\eta^2} - \frac{2B_0^2 \eta_0^4 (\o_r^2 + \k)}{\h^4} = 0,
\ee
with solutions
\be
\o_s = \frac{1}{8\omega_r} \left[ \left(\k -\frac{2}{\eta^{2}}\right) \pm \frac2{\h^2} \sqrt{1 + 4B_0^2\eta_0^4(\o_r^2+\k) + 3\kappa\eta^2 + \frac{9}{4}\kappa^2\eta^4} \right].
\label{quad_dust}
\ee

We point out three particularly interesting limits in this expression.  In the limit of zero curvature $\kappa\rightarrow0$: 
\begin{equation}
\o_s = -\frac{1}{4\omega_r\eta^2} \left[ 1 \pm \sqrt{1 + 4B_0^2 \eta_0^4\o_r^2} \right]~;
\end{equation}
in the  high-frequency $\omega_r^2 \gg 1$ where we once again recover the static solution:
\begin{equation}
\frac{1}{L} \equiv \o_s = \frac{B_0}{2 a}~;
\end{equation}
and at extremely short wavelengths $(k\eta) \gg 1$:
\be
\o_s = \frac{\k}{2\o_r}\ , \ \frac{-\k}{4\o_r} ~, 
\ee
where we  recover both earlier solutions (\ref{wk1})  obtained by dropping all high order terms.

We thus see that the Gertsenshtein effect is operable in this regime, but that the time-dependent solution damps out at an even faster rate than in radiation-dominated models.\\

We  should also make a note concerning the presence of charged particles in RD and MD models.  In the real universe, recombination occurs at $z \sim 10^3$, when the universe is already matter dominated.  Above recombination temperatures, one would not expect the Gertsenshtein mechanism to operate in the standard RD model due to Thomson scattering of photons off free electrons, which makes the universe opaque to electromagnetic radiation.  Below recombination temperatures and above reionization (the "dark ages") one can model the presence of charged particles in a somewhat {\it ad hoc} fashion similarly to the procedure in Domcke and Garcia-Cely \cite{DCG21} by adding a term $(\o_{p0}^2/a) b$ (in conformal time) to the LHS of the second of the wave eqs. (\ref{bheqn}).  Here, $\o_{p0} \equiv \sqrt{e^2 n_{e0}/m_e}$ is today's value of the plasma frequency. For the current value of $B_0 \lesssim 10^{-9}$ G, our requirement that $B_0^2 \ll \r_0$ for a RD or MD universe is easily met, as is the constraint that $k\h \gg 1$ for EMWs.  Then, repeating the above derivation for the MD case, the solution becomes
\be
\o_s = \frac{1}{8\omega_r} \left[ \k -\frac{2}{\eta^{2}} + \overline\o_p^2 \pm
\frac2{\h^2} \sqrt{ 1 + 4B_0^2\eta_0^4(\o_r^2+\k) + 3\kappa\eta^2 + \frac{9}{4}\kappa^2\eta^4 + \frac32\overline\o_p^2\kappa\eta^4 + 
\overline\o_p^2\eta^2 + \frac{\overline\o_p^4\eta^4}{4}} ~\right].
\label{quad_plasma}
\ee
where $\overline\o_p^2 \equiv \o_{p0}^2/a = \o_{p0}^2\h_0^2/\h^2$.

In the case that spatial curvature is negligible, and ignoring all but the frequency-dependent terms appropriate for the high frequency limit, (\ref{quad_plasma} reduces to
\be
\o_s = \frac{\overline\o^2_p}{8\o_r} \left[ ~1 \pm 
\sqrt{1 + \frac{16B_0^2\o_r^2}{a^2\overline\o^4_p}} ~\right] . 
\label{o_pMDsol}
\ee
When the second term under the radical dominates, this expression becomes our previous result $\omega_s = {B_0}/{2 a}$.  For $\o_{p0} \sim 1$ Hz, as in \cite{DCG21}, that  takes place  when $\o_r \gtrsim 10$ MeV.  Below these energies effects of the plasma need to be taken into account.
We see that when $\o^4_{p0} \gg 16B_0^2\o_r^2$ with the (+) root, the above reduces to 
\be
L^{-1} = \o_s = \frac{\o_{p0}^2 }{4 a \o_{r}} = \frac{\o_{p0}^2 }{4\o_{r0}},
\label{o_pMDsol_L}
\ee
independent of redshift, since $\o_r = \o_{r0}/a$.  When evaluated in terms of coordinate time, this expression agrees with the one given in  \cite{DCG21}, $\o_s \sim (1+z)^2 \o_{p0}^2/4\o_{r0}$.

However, this result should not be interpreted as the mixing length associated with the Gertsenshtein effect: Since $B_0$ no longer appears in (\ref{o_pMDsol_L}),  graviton-photon oscillations cannot take place. Instead, it gives a ``coherence length" or ``skin depth" above which the EMW does not propagate.  On the other hand, after expanding the radical in (\ref{o_pMDsol}) when the plasma frequency dominates, the negative root provides a rough scaling of the Gertsenshtein effect $\o_s \propto B_0^2 \o_r / (a\o_{p})^2$, which has the anticipated behavior that $\o_s \rightarrow 0$ when $B_0\rightarrow0$, $\o_p\rightarrow\infty$, and $a\rightarrow\infty$. We note that the inclusion of electric charges produces results similar to those pointed out by Zel’dovich in his original objections to the Gertsenshtein effect \cite{Zel73}.

\section{Curvature-dominated universe}\label{CD}

As discussed in \S\ref{subRD}, the RD and MD cases have a limited domain of validity, after which the model becomes curvature dominated. For the CD regime, the Friedmann equation in conformal time (\ref{eqn:friedman}) becomes simply
\be
\left(\frac{\dot a}{a}\right)^2 = -\k, \label{CDFriedmann}
\ee
for $\k = \pm 1$. Here we must assume the constraint $B_0^2 \ll |\k| a^2$. Curvature-dominance simplifies  the GW equation (\ref{eqn:nonflat_h} considerably by replacing a time-dependent expansion rate in the damping term by the constant $H\equiv \dot{a}/a$. Perhaps not surprisingly, equations (\ref{eqn:nonflat_h}) and (\ref{eqn:nonflat_b}) can then be satisfied by the same ansatz (\ref{dSansatz}) we employed in the de Sitter case, which is  applicable to both positive and negative curvature.

Substituting the expressions (\ref{dSansatz}) into (\ref{eqn:nonflat_h}) and (\ref{eqn:nonflat_b}), we find from the GW equation
\begin{equation}
{\cal A} \left[-4\omega_r\omega_s + 2i(\omega_r + \omega_s)(\g - H)
             -2H\g  +\gamma^2 - \kappa \right] = - {\cal B} \left( 4 B_0 a^{-2} e^{\gamma\eta} \right) ~,
\label{eqn:dispersion_h}
\end{equation}
and from the EMW equation
\begin{equation}
{\cal A} B_0 \left[(\omega_r - \omega_s)^2 + \kappa\right]
             = - {\cal B} \left(2\kappa - 4\omega_r\omega_s \right) e^{\gamma\eta} ~.
             \label{eqn:dispersion_b}
\end{equation}
Again, these expressions are valid for both positive and negative curvature spacetimes, which in the
regime  considered here differ only by the value of $H$.

At this point the positive and negative curvature solutions take different forms, depending on whether $H$ is real or imaginary (the latter will be true in the closed $\k = +1$ topology  with its oscillatory behavior.)  For $\k =-1$, requiring that the imaginary parts of equation (\ref{eqn:dispersion_h})  vanish gives $\gamma=H=1$.  Then solving (\ref{eqn:dispersion_h}) and (\ref{eqn:dispersion_b}) together to eliminate $\cal{A}/\cal B$ gives
\begin{equation}
\frac{(1+2\o_r\o_s)}{(\o_r-\o_s)^2 - 1} = \frac{B_0^2}{2a^2 \o_r\o_s}.
\end{equation}
This is a quadratic equation for $\o_s$ and with $(\o_r + \o_s)^2 \approx \o_r^2$, yields
\be 
\o_s = \frac1{4\o_r}\left[-1 + \sqrt{1 + \frac{4B_0^2(\o_r^2 -1)}{a^2}}\right] ~. \label{osk-1}
\ee 
We have chosen the (+) root in (\ref{osk-1}) because $B_0 = 0$ in (\ref{eqn:dispersion_h}) implies $\o_s = 0$. When the second term in the discriminant dominates and $\o_r \gg 1$ and $a = 1$, this result reduces to the static case (\ref{Lstat}). The condition $\o_r \gg 1$ is, inserting dimensions, really $\o_r \gg H$; thus $\o_r \gg 1$  is satisfied for any wave whose period is much shorter than $H^{-1}$. Moreover, the utility of the plane-wave ansatz evidently  assumes that $a\sim$ constant, in which case the GW equation becomes essentially that of a damped harmonic oscillator.  Consequently, $\o_r \gg 1$ should hold in any remotely plausible scenario.  We also see, however, that $a = 1$ gives the maximum value of $\o_s$ and as the universe expands, $\o_s$ decreases monotonically to zero, quenching photon-graviton oscillations.\\ 

For the $\k = +1$ case, $H= \pm i$. In order for the imaginary part of (\ref{eqn:dispersion_h}) to vanish, one must take $\g = 0$, leaving
\be
{\cal A} \left[-4\omega_r\omega_s  - 1 \pm 2(\omega_r + \omega_s)\right] = - {\cal B} \left( 4 B_0 a^{-2} e^{\gamma\eta} \right).\label{dishHi}
\ee
Similarly to above, with (\ref{eqn:dispersion_b}) this gives for $H = + i$
\be
\frac{(4\o_r\o_s -2)}{(\o_r - \o_s)^2+1} = \frac{4B_0^2}{a^2[4\o_r\o_s - 2(\o_r+\o_s)+1]}.
\ee
Assuming again that $\o_r \gg \o_s$, we get  the quadratic dispersion relationship
\be
16\o_r^2\o_s^2 - 4\o_r(2\o_r+1)\o_s + 2(2\o_r -1) - \frac{4B_0^2(\o_r^2 + 1)}{a^2} =0,
\ee
which can be solved for $\o_s$ to get
\be
\o_s = \frac1{8\o_r}\left[1 + 2\o_r \pm \sqrt{4\o_r^2 -12\o_r + 9+ \frac{16B_0^2(\o_r^2 + 1)}{a^2}}\right].\label{k1sol}
\ee

For $B_0= 0$, (\ref{dishHi}) requires $\o_s = 1/2$.  This is obtained with the positive root  in (\ref{k1sol}) and the previous assumption that $\o_r \gg 1$. The static limit is recovered when the last term in the discriminant dominates and $a \sim 1$, but more generally we have for the  mixing length
\be
L = \frac{1}{\omega_s} = \frac{2a}{B_0}.
\ee
Note that despite the presence of $a$ in all our  expressions for $L$, they represent comoving lengths, since $\o_s = k_s$ comes from $e^{-i k_s(z+\h)}$, where $z$ is a comoving coordinate. 

\section{Discussion}
\label{Discussion}

We have shown that in radiation- and matter-dominated models, the Gertsenshtein mechanism can operate in the limit $k\h \gg 1$, that is at times and spatial scales when the  graviton or photon wavelength is much shorter than the Hubble radius.   For curvature-dominated models, we found mixing solutions for both positive and negative curvature, assuming only that the expansion rate is slow compared to the particle frequency. Nevertheless, in all these cases it appears that  oscillations  damp out as the universe expands. For the de Sitter universe we found similar behavior: the Gertsenshtein effect can take place only at early times ($\eta \ll H^{-1}$) and for perturbation wavelengths much smaller than the Hubble radius. In all cases, for superhorizon wavelengths, the perturbations freeze in and oscillations cease.

 When the effect of charged particles are included (during the ``dark ages") we find that charge density oscillations at the plasma frequency impose a strong constraint on photon-graviton mixing, essentially suppressing mixing at EMW frequencies below $\sim 10$ MeV, and, as discussed earlier, causes the mixing frequency to decrease with the square of the plasma frequency (within the coherence length of Gertsenshtein mixing). Such effects are not important for a curvature-dominated universe or a cosmological-constant--dominated universe, where coupling to GWs occur through the spacetime geometry, not its matter content.  For radiation-dominated models we do not expect the Gerstenshtein effect to operate at all, due to Thomson scattering of EMWs off electrons. 
 
 Turning to  regimes where the GE operates without plasma dissipation, and given that today's intergalactic magnetic field $\lesssim 10^{-9}$ G implies that the mixing length for graviton-photon conversion is $\gtrsim 10^5$ Hubble radii, we should confront the question raised at the start of this paper: Does a ``cosmological" Gertsenshtein effect have any observable consequences?

 Initially, one might hope that it does. Presumably, the cosmological magnetic field was  much larger in the early universe than it is today, and the Gertsenshtein effect (or similar mixing process, such as   axion-photon mixing or graviton-Yang-Mills-boson mixing) could  conceivably provide a  mechanism for altering the balance of particle species during an inflationary phase.
 
 Our current assessment is, however, that the outlook appears not to be bright.  In conformal time the Hubble radius is $H^{-1} = a^{2}/\dot a$.  The  proper mixing length, for all models and regimes where the GE operates unhindered, is $L_p = aL = 2a^2/B_0$ for constant $B_0$. Notice, however, that this can be written as $L_p = 2/|B|$, where $|B| = B_0/a^2$ is the magnitude of the background field when $B$ is defined covariantly as in (\ref{eqn:pertB}).  Thus $L_p$ evidently gives an ``invariant" measure of the mixing length, which can be applied to both the static or expanding cases.  Hence in general
\be
\frac{L_p}{H^{-1}} = \frac{2 \dot a}{B_0}.
\ee
For the RD and MD cases respectively, the Friedmann equation (\ref{eqn:friedman}) then shows 
 \be
\frac{L_p}{H^{-1}} \approx \frac{\sqrt{\r_0}}{B_0} \gg 1 \quad ; \quad \frac{L_p}{H^{-1}} \approx \frac{ \sqrt{a \r_0}}{B_0} \gg 1.
\ee
For de Sitter, the horizon is not the Hubble radius, but the parameter in our results is $H$ and one easily finds that $L_p/H^{-1} \gg 1$ as well. 
 
Consequently, photon-graviton oscillations always lie outside the Hubble radius and are in principle unobservable.  The same would be the case for the SU(2) boson-graviton oscillations investigated in \cite{PR23b}.  There, the mixing length was found to be essentially the same, with $B_0$ replaced by $qA^2$, for coupling constant $q$ and potential $A$, such that $q^2A^4 \ll 1$.  We do point out that the lack of observability also holds for superadiabatic amplification of gravitational and electromagnetic waves \cite{Tsagas:2004kv, Grish74}, which also takes place at superhorizon scales.

One might suspect  a possible ``out" is to consider a scalar field $\f$ instead of a magnetic field $b$ in a potential-dominated inflationary phase.  In that case our isotropy constraint might be evaded because the (isotropic) scalar potential $V$ would replace $B_0^2/a^2$ and also drive the expansion such that $H \sim V^{1/2}a$ (see \ref{eqn:friedman}). Then $L_p$ could become as short as $\sim H^{-1}$, which is in fact what one expects for quantum mechanically excited modes in de Sitter \cite{TW88}.  In such a scenario, the Gertsenshtein mechanism could have an observable effect on the particle make-up of the universe.

Unfortunately,  coupling gravitons to scalar modes in this scenario is nontrivial.  Replacing $T_{\m\n}^{EM}$ in the Einstein equations with the standard scalar-field stress-energy tensor $T_{\m\n} = \f,_\m\f,_\n - [V(\f) + \frac12\f^{,\a}\f,_\a] g_{\m\n}$ and assuming potential dominance, yields for both the $h_{11}$ and $h_{12}$ modes exactly (\ref{eqn:nonflat_h}) with a zero RHS.  In other words, at first order there is no coupling between $h$ and $\f$.

In short, although a Gertsenshtein-like mechanism could in principle affect the particle balance in an inflationary period by catalyzing inter-species conversion, in practice it is ineffective unless a stronger coupling among fields can be engineered.  Whether that is best carried out in an $f(R)$ or other higher-order theory of gravity, or in an anisotropic cosmology,  remains the subject for future investigation.

\vskip19pt
\leftline{\large\bf Acknowledgements}

The work by P.A. was performed under the auspices of the U.S. Department of Energy by Lawrence Livermore National Laboratory under contract DE-AC52-07NA27344.


\end{document}